\documentclass{article}

\usepackage{arxiv}

\usepackage[utf8]{inputenc} 
\usepackage[T1]{fontenc}    
\usepackage{hyperref}       
\usepackage{url}            
\usepackage{booktabs}       
\usepackage{amsfonts}       
\usepackage{nicefrac}       
\usepackage{microtype}      
\usepackage{cleveref}       
\usepackage{lipsum}         
\usepackage{graphicx}
\usepackage{natbib}
\usepackage{doi}

\title{Recursive Detection and Analysis of Nanoparticles in Scanning Electron Microscopy Images}

\author{ \href{https://orcid.org/0000-0001-6877-1328}{\includegraphics[scale=0.06]{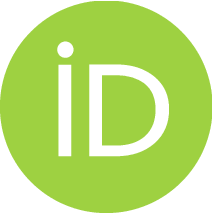}\hspace{1mm}Aidan S.~Wright}
\\
	Department of Engineering\\
	Oral Roberts University\\
	Tulsa, OK 74171 \\
	\texttt{aidan\_w@oru.edu} \\
	\And
	\href{https://orcid.org/0009-0009-9064-5606}{\includegraphics[scale=0.06]{orcid.eps}\hspace{1mm}Nathaniel P. ~Youmans} \\
	Department of Computing and Mathematics\\
	Oral Roberts University\\
	Tulsa, OK 74171 \\
	\texttt{nyoumans@oru.edu} \\
	\And
	\href{https://orcid.org/0000-0001-5897-6573}{\includegraphics[scale=0.06]{orcid.eps}\hspace{1mm}Enrique F.~Valderrama Araya, Ph.D.} \\
	Department of Computing and Mathematics\\
	Oral Roberts University\\
	Tulsa, OK 74171 \\
	\texttt{evalderrama@oru.edu} 
}

\renewcommand{\shorttitle}{Recursive Detection \& Analysis of Nanoparticles in SEM Images}

\begin{document}
\maketitle

\begin{abstract}
	In this study, we present a computational framework tailored for the precise detection and comprehensive analysis of nanoparticles within scanning electron microscopy (SEM) images. The primary objective of this framework revolves around the accurate localization of nanoparticle coordinates, accompanied by secondary objectives encompassing the extraction of pertinent morphological attributes including area, orientation, brightness, and length.

Constructed leveraging the robust image processing capabilities of Python, particularly harnessing libraries such as OpenCV, SciPy, and Scikit-Image, the framework employs an amalgamation of techniques, including thresholding, dilating, and eroding, to enhance the fidelity of image processing outcomes.

The ensuing nanoparticle data is seamlessly integrated into the RStudio environment to facilitate meticulous post-processing analysis. This encompasses a comprehensive evaluation of model accuracy, discernment of feature distribution patterns, and the identification of intricate particle arrangements. The finalized framework exhibits high nanoparticle identification within the primary sample image and boasts  97\% accuracy in detecting particles across five distinct test images drawn from a SEM nanoparticle dataset (\cite{aversa_first_2018}). Furthermore, the framework demonstrates the capability to discern nanoparticles of faint intensity, eluding manual labeling within the control group.
\end{abstract}


\section{Introduction}

Scientific research generates a multitude of SEM (Scanning Electron Microscope) images of particles, which often necessitate manual labeling and data extraction. Disciplines like biology and physics extensively employ imaging techniques to analyze micro/nano features on surfaces (\cite{nanou_scanning_2018}). This procedure is characterized by its slow pace, labor-intensive nature, and susceptibility to human errors. Enabling computers to autonomously recognize and characterize features within images from these fields is crucial for large-scale data analysis. Frequently, a dataset requires manual labeling to furnish a machine learning model with training data. An alternative to manual labeling involves utilizing semi-synthetic datasets to train neural networks (\cite{lopez_gutierrez_nanoparticle_2022}). Once trained, such models can accurately label other images. For instance, this technology allows biologists to determine cell counts within an image, often with higher precision than manual methods. These applications extend beyond particle recognition, encompassing tasks like crack detection (\cite{zhao_hybrid_2021}).\\

An intriguing application of nanoscale image feature recognition lies in defect detection in silicon wafers (\cite{kim_automatic_2017}). Trained neural networks can go beyond image analysis by predicting nanoparticle distributions based on background characteristics (\cite{ayush_nanonet_2022}). Various open-source computer programs exist for particle recognition and characterization, with ImageJ being a notable example. ImageJ, a public domain software, specializes in scientific image analysis and has been employed to label images for subsequent use in convolutional neural network training (\cite{schneider_nih_2012}).\\

ImageJ allows users to manually input and process images, rendering them amenable to analysis and data recording. However, one drawback of ImageJ is its reliance on users to select appropriate image processing techniques for particle analysis. As ImageJ lacks specificity for nanoparticle recognition, users must familiarize themselves with its graphical user interface, potentially posing challenges for those unacquainted with image processing. The model presented in this study draws inspiration from various image processing techniques employed in ImageJ but establishes a streamlined process whereby a computer autonomously analyzes multiple images in a folder and collects data without requiring user intervention. This program facilitates particle data extraction from SEM images for any user, regardless of their familiarity with image processing techniques.

\section{Methods}
The model was developed using Python, primarily utilizing OpenCV—an established library tailored for computer vision applications. Throughout the project, OpenCV served as the core library. Supplementary libraries, including SciPy, Pillow, and Scikit-Image, were also incorporated. SciPy, a comprehensive library with wide-ranging applications in scientific contexts like data processing and system prototyping, played a significant role. Pillow, a derivative of the Python Imaging Library, contributed additional image processing functions. Similarly, Scikit-Image, akin to OpenCV, provided capabilities in image segmentation and feature detection. These integrated libraries collectively enriched the project's capabilities and utility.

\begin{figure}
	\centering
	\includegraphics[width=\textwidth]{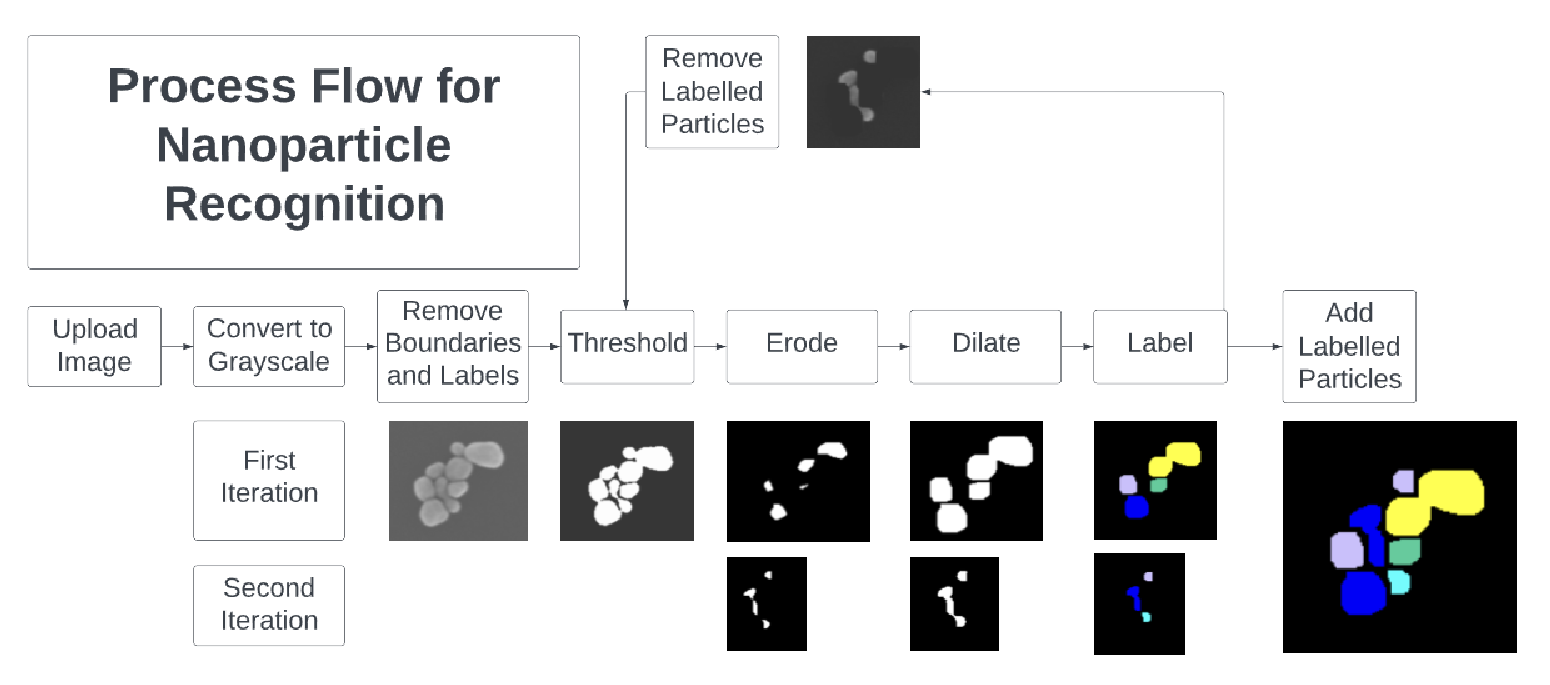}
	\caption{Process flow for nanoparticle recognition.}
	\label{fig:fig1}
\end{figure}

The process flow for nanoparticle recognition using the recursive pipeline is illustrated in Figure \ref{fig:fig1}. The procedure begins with image uploading, followed by grayscale conversion, where pixel values of 0 represent pure black and pixel values of 255 represent pure white. Subsequent stages of the process are also detailed in Figure \ref{fig:fig1}. Initial thresholding divides the image into pure black and pure white segments. This is followed by erosion, which reduces particle size, and dilation, which enlarges features. Subsequently, contiguous clusters of white pixels are identified and labeled as nanoparticles. The recursive nature of the process comes into play once the initial labeling is complete. The labeled regions are replaced with the background's pixel value in the original image. This effectively masks the stronger pixels, enabling the thresholding process to spotlight nanoparticles that might not have met the threshold initially. These newly detected particles undergo the same sequence of thresholding, erosion, dilation, and labeling before being incorporated into the final set of labeled nanoparticles.\\

In the image processing model's first step, thresholding was employed, utilized both in the Hough transform and ImageJ. The objective was to convert the background to pure black and the foreground (nanoparticles) to pure white. However, a challenge arose regarding the appropriate threshold level within the grayscale pixel values ranging from 0 to 255, where 0 signifies pure black and 255 signifies pure white. If the threshold was set too high, only the brightest nanoparticles would be distinguishable, potentially classifying darker particles as part of the black background. Given that lighting and contrast varied across images, adopting a fixed threshold for each image would introduce issues of reproducibility and parameterization, akin to the challenges encountered in the preceding image analysis approach.

\begin{figure}
	\centering
	\includegraphics[width=0.4\textwidth]{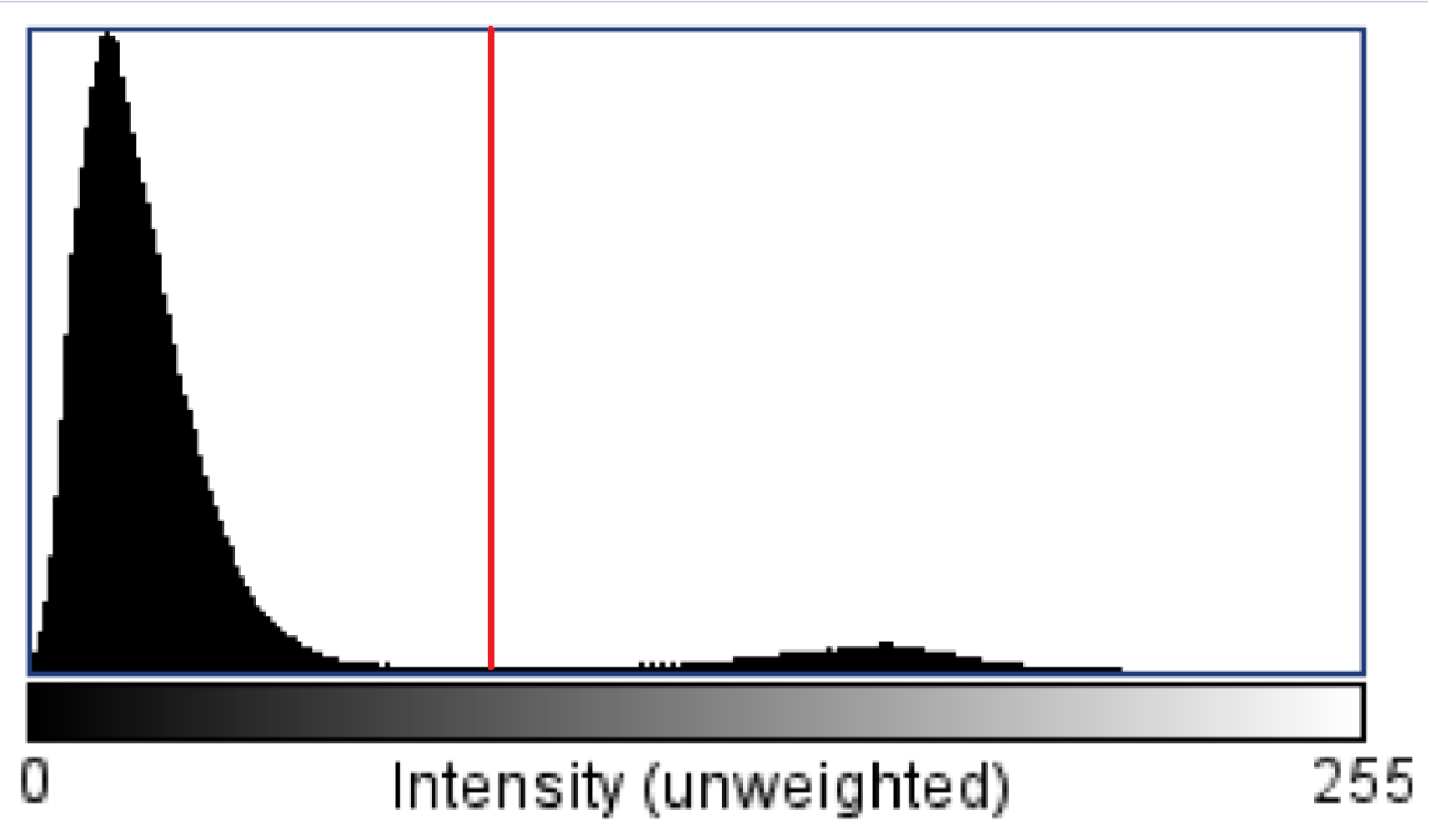}
	\caption{Intensity plot of each pixel value in a sample SEM nanoparticle image.}
	\label{fig:fig2}
\end{figure}

\begin{figure}
	\centering
	\includegraphics[width=\textwidth]{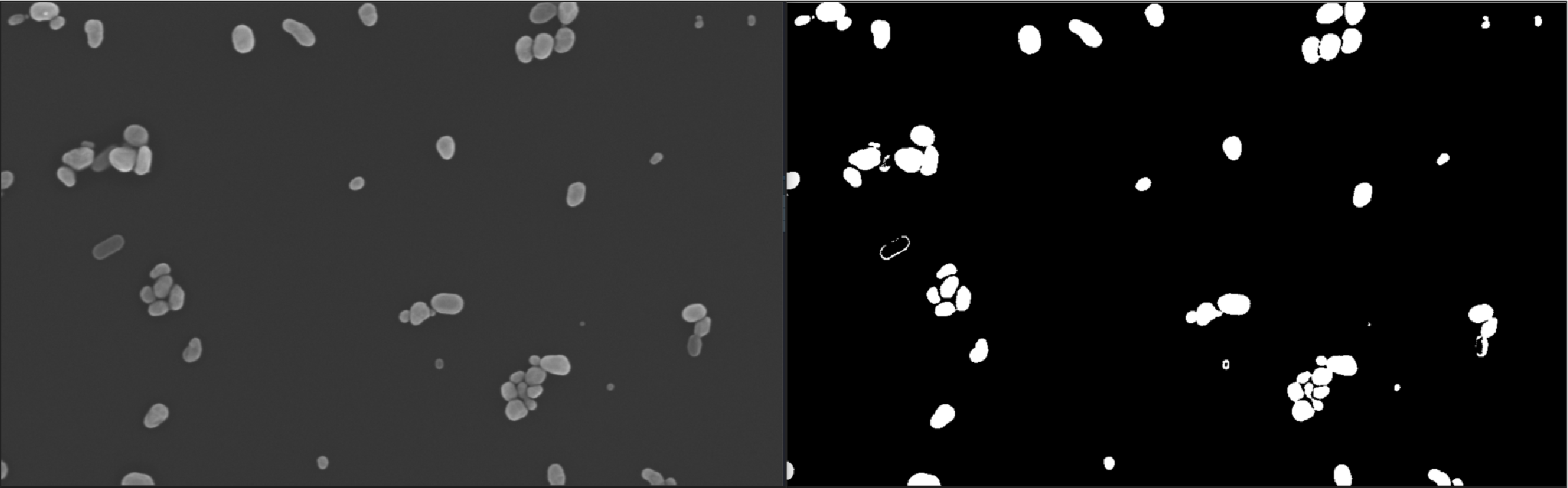}
	\caption{Example image containing nanoparticles, where a threshold method(Otsu's) allows the identification and position of individual nanoparticles.}
	\label{fig:fig3}
\end{figure}

Figure \ref{fig:fig2} depicts a plot showcasing the pixel intensity distribution within a sample SEM nanoparticle image. Notably, a pronounced concentration of particles is observed within the 0 to 60 range, indicative of the predominantly black background. A subtle elevation in pixel counts between the 117 to 186 range signals the presence of a distinct cluster of nanoparticles standing out against the background. Insight into the ranges encompassing these two groups is instrumental in selecting an optimal threshold value to effectively distinguish foreground from background. To address this, an established image processing technique known as thresholding is employed, and specifically, Otsu's method is introduced. This algorithm, designed to fine-tune threshold determination, automatically computes an image's threshold based on inherent pixel value distributions as depicted in the graph. When applied to images containing nanoparticles, Otsu's method effectively isolates and highlights nanoparticles within the image, as exemplified in the Figure \ref{fig:fig3}. Despite its efficacy, this algorithm occasionally falls short in capturing all nanoparticles within the thresholding process. Consequently, a subsequent stage is dedicated to processing these particles individually, ensuring comprehensive analysis.

A matrix composed of 1s and 0s, corresponding to white and black pixels, respectively, is generated based on the processed binary image. Subsequently, this mask is superimposed onto the original image, and the label function provided by SciPy is employed to create an array encapsulating each distinct object. While this methodology effectively labeled white pixels within the image, it fell short in distinguishing between grouped nanoparticles and individual ones, leading to the labeling of multiple particles as a single entity. Moreover, the thresholding process, though commendable, is not infallible. Any residual background noise that surpasses the thresholding stage is misidentified as nanoparticles, and nanoparticles only partially converted to white are left incompletely labeled. To address this, a suite of image processing functions from OpenCV is employed to refine the image. The erode function, a key element of OpenCV, facilitates the reduction of nanoparticle dimensions by eliminating white pixels along the black pixel borders. This serves to effectively counter noise, which tends to possess an abundance of black pixels susceptible to shrinkage. This operation also resolves the issue of inadequately labeled particles, a concern resolved in subsequent steps. \\
To accurately restore the actual nanoparticle sizes, black pixels that are in contact with white pixels are substituted with white pixels. The dilation process is executed using OpenCV's dilate function. By augmenting nanoparticle dimensions by the same number of pixels they were eroded by, the particles, for the most part, regain their original size. This procedure additionally aids in separating particles that were in slight contact.

\begin{figure}
	\centering
	\includegraphics[width=0.7\textwidth]{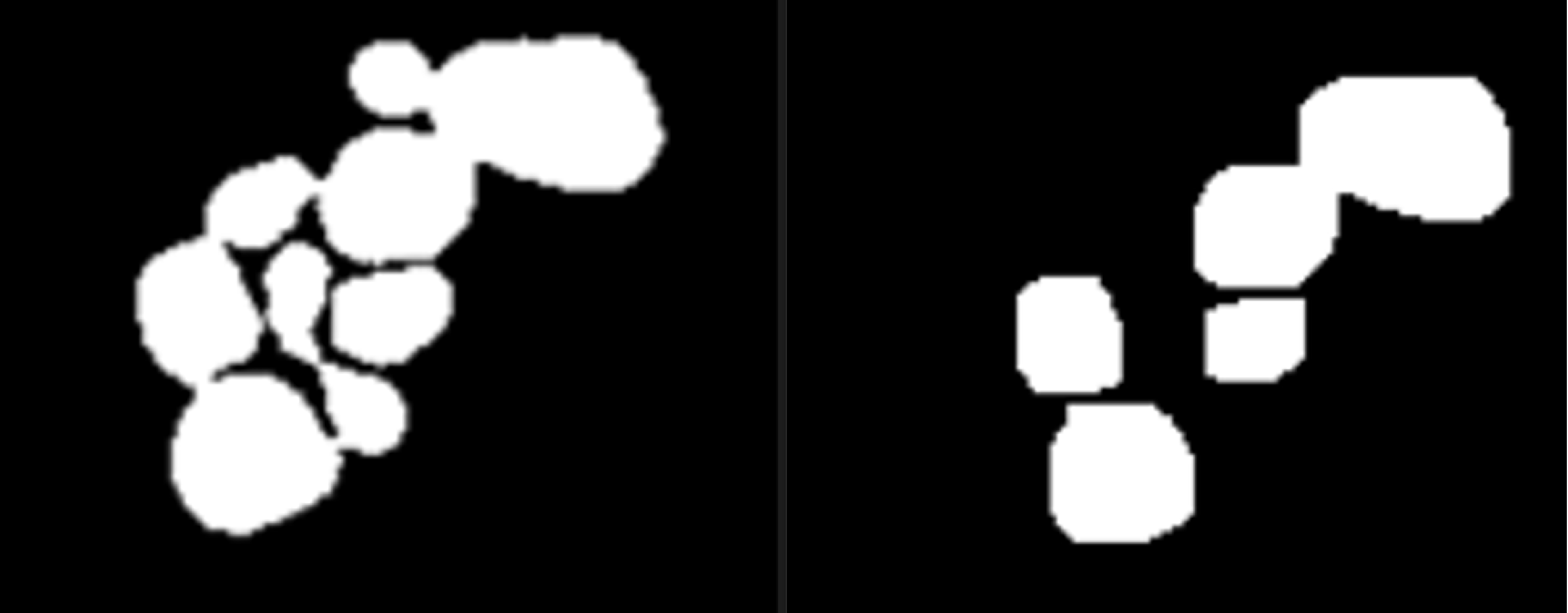}
	\caption{Separation of particles achieved by eroding and dilating the image. The smaller nanoparticles can be removed and recovered in the next iteration.}
	\label{fig:fig4}
\end{figure}

\begin{figure}
	\centering
	\includegraphics[width=0.8\textwidth]{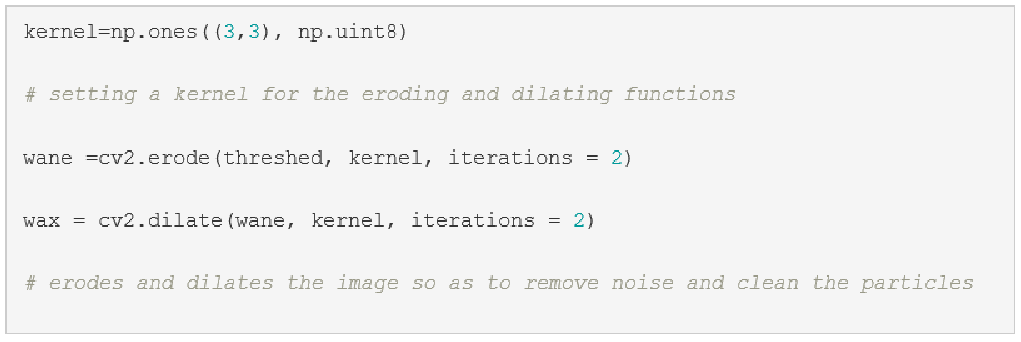}
	\caption{Python script to separate particles using erode/dilate functions from the OpenCV library.}
	\label{fig:fig5}
\end{figure}

The particle separation process is demonstrated in Figure \ref{fig:fig4}, where all nanoparticles appear to be in contact except for one outlier. The initial image's left side would have led the labeling function to categorize the cluster of particles as merely two entities. Employing an erosion and dilation sequence, the image is processed to eliminate smaller nanoparticles (which are subsequently recoverable) and isolate the larger particles, as evident on the image's right side. It's important to note that this straightforward image processing technique isn't flawless, as illustrated by the uppermost two particles in the right-side image. Despite its imperfections, the approach succeeds in the separation of the majority of particles. The provided Code 1 (Figure \ref{fig:fig5}) outlines the Python implementation of this process.

\begin{figure}
	\centering
	\includegraphics[width=\textwidth]{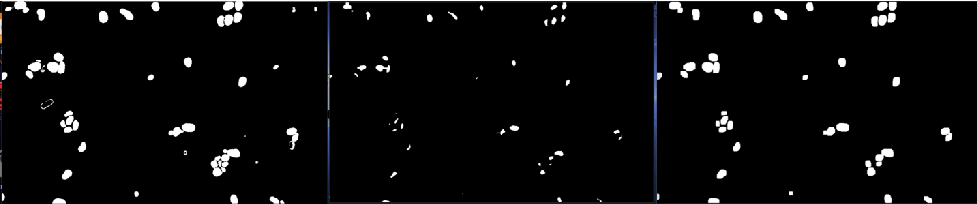}
	\caption{The eroding and dilating process by comparing the raw thresholded image on the left, the eroded image in the middle, and the dilated image on the right.}
	\label{fig:fig6}
\end{figure}

Figure \ref{fig:fig6} provides a visual representation of the erosion and dilation process, showcasing the unprocessed thresholded image on the left, the eroded iteration in the center, and the dilated rendition on the right. This methodical progression enhances the precision of labeling in the final image. \\

The subsequent concern pertains to the exclusion of smaller, dimmer particles that undergo eradication during the erosion phase. These nanoparticles, which elude complete manifestation in the initial processing due to their partial emergence during thresholding and subsequent removal in the erosion process, warrant specific consideration.\\

\begin{figure}
	\centering
	\includegraphics[width=\textwidth]{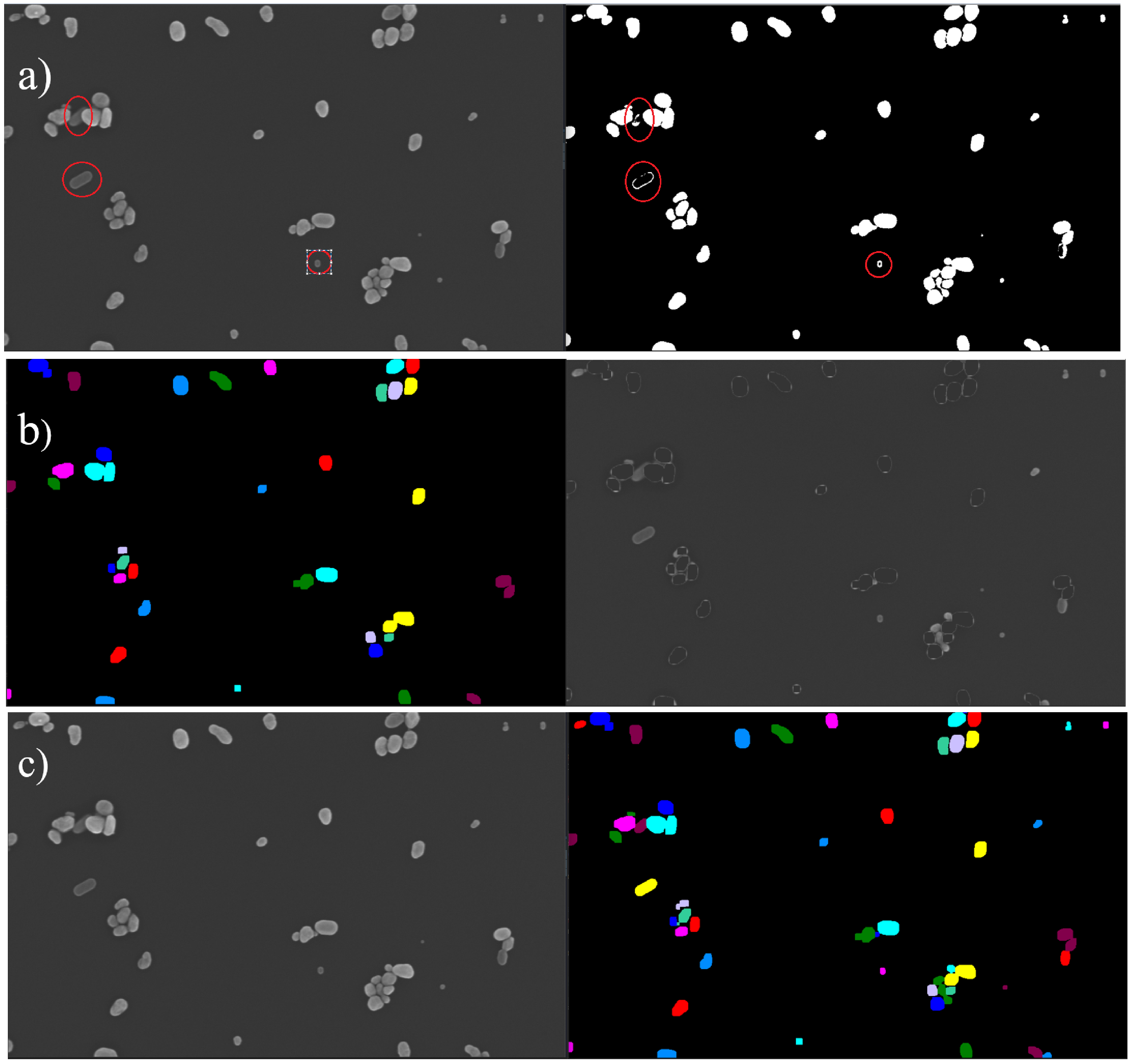}
	\caption{Example of how an iteration of thresholding, eroding and dilation process works to increase the amount of nanoparticles detected.}
	\label{fig:fig7}
\end{figure}

Figure \ref{fig:fig7} serves as an illustrative example of how to enhance the detection of nanoparticles through an iterative process.

In Figure \ref{fig:fig7}a, nanoparticles from the SEM image (left) are highlighted. These nanoparticles would not be detected by applying only thresholding.

In Figure \ref{fig:fig7}b, the left side shows nanoparticles detected in the first iteration (colored), while the right side displays the new image generated after removing the nanoparticles identified in the previous step. The pixel intensities at the coordinates of the first-iteration nanoparticles are replaced with the mean intensity of the image, effectively "removing" the easily identifiable nanoparticles. This results in an image with fainter structures, where the difference between the maximum and minimum intensities is reduced. Subsequently, a new lower thresholding is applied to this modified image, followed by a gentler erosion and dilation process, which enables the detection and classification of the fainter nanoparticles.

In Figure \ref{fig:fig7}c, a comparison is shown between the original SEM image (left) and the colored and labeled image on the right, obtained after a second round of thresholding, erosion, and dilation steps.


Once the particles were fully detected, the Python library ScikitImage was used to measure the properties of each detected particle, such as the location, size, orientation, and brightness levels. These results were exported to a csv file which was statistically analyzed using R.

\section{Results}

\begin{figure}
	\centering
	\includegraphics[width=0.8\textwidth]{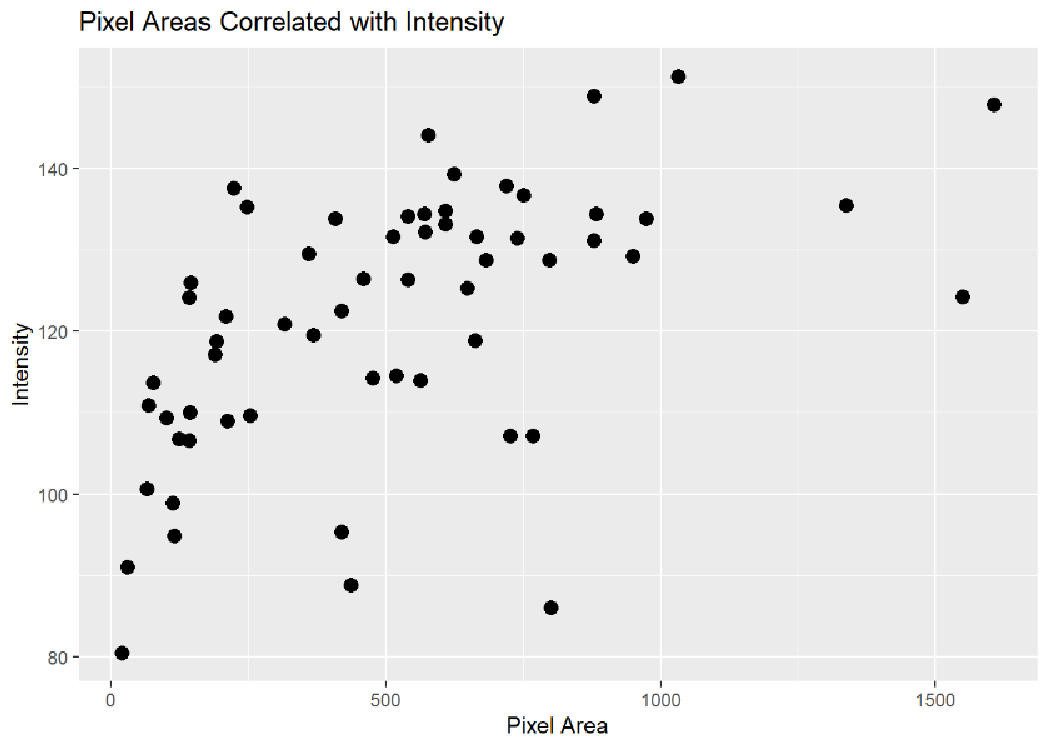}
	\caption{Correlation between intensity and size of the nanoparticles.}
	\label{fig:fig8}
\end{figure}

The primary results from this model were the location and area of the nanoparticles within the image. The file contained the X coordinates and the Y coordinates (in pixels) of each nanoparticle, the size of each particle (in pixels), major axis, minor axis, perimeter, and mean intensity. This information was used to do some basic analysis of the particles. Figure \ref{fig:fig8}, represents the correlation between intensity and size of the nanoparticles.

\begin{figure}
	\centering
	\includegraphics[width=0.8\textwidth]{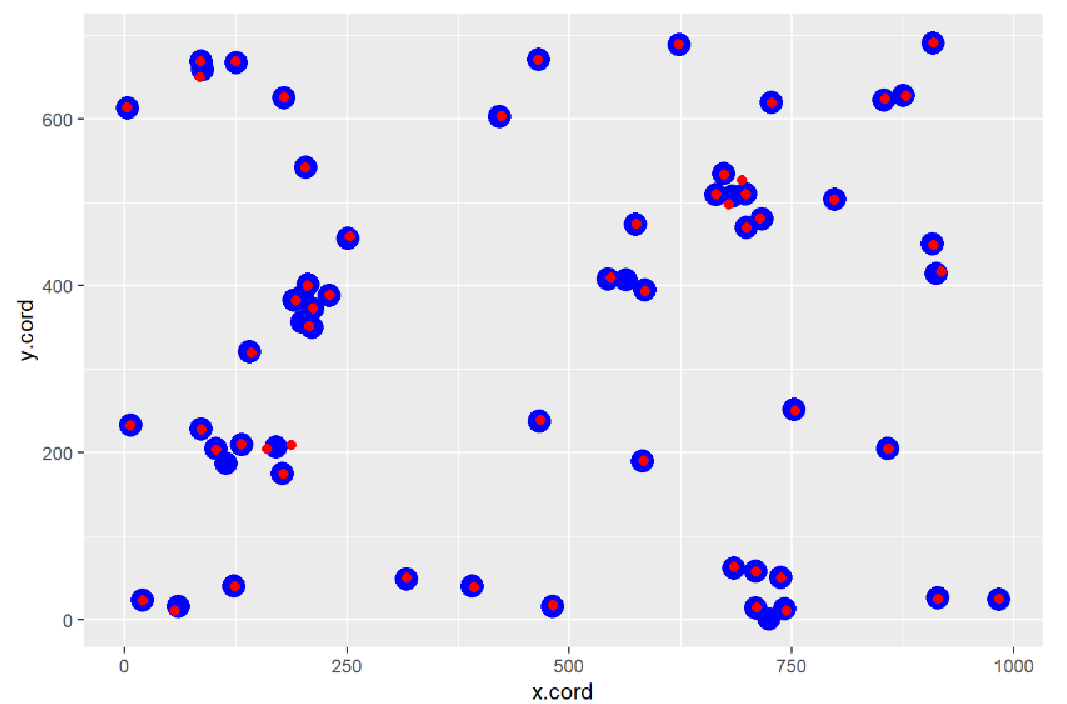}
	\caption{Comparison of computer labeled nanoparticles (blue) and human labeling (red). The iteration find three more than by simple eyesight. }
	\label{fig:fig9}
\end{figure}

The most important outcome from the analysis in R was the accuracy of the nanoparticle detection. Figure \ref{fig:fig9} is a graph of an example nanoparticle image and shows the computer modeled locations of nanoparticles (blue) versus the hand labeled nanoparticle locations (red).

Figure \ref{fig:fig9} provides a lot of useful information as it shows where the model surpassed human labeling and where it failed to distinguish between particles. This image had a 96.5\% accuracy rate of matching human labeled nanoparticles but found three nanoparticles that were originally unnoticed by human labeling.\\

As an added application, the model tested outside of the original SEM image application and was tested on images of grains of sand. Sand proved to be a much trickier particle to separate and detect than the SEM nanoparticles. The nanoparticles imaged tended to be circular and have little variation in grayscale value within each particle. In contrast, sand comes in several different shapes, with the more manageable shapes being squares, triangles and thin rectangles, and the more extreme shapes resembling Y or doughnut structures. This shapes proved to be more difficult to recognize as it was harder to determine how much each image should be eroded or thresholded. For example, Y shapes had more white pixels in contact with black pixels than a similar sized particle with a circular shape. This causes Y and doughnut shapes to erode much faster than circular or square shapes. Additionally, Y shapes are prone to being separated into multiple different pieces using water-shedding. 
The biggest challenge when applying our model to an image of sand, was the large variation in gray-scale value within each particle.

\section{Discussion}

\begin{figure}[h!]
	\centering
	\includegraphics[width=0.7\textwidth]{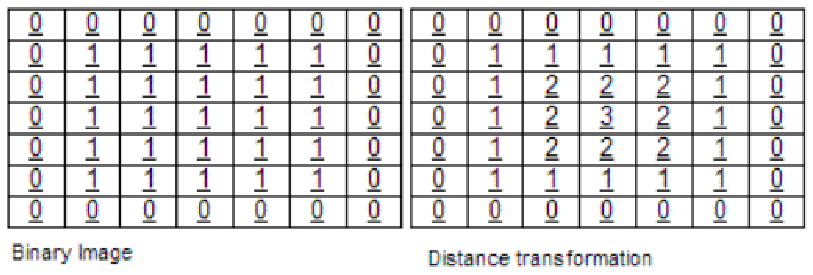}
	\caption{Comparison between matrix of a binary image with the distance transformation matrix, which illustrate how water-shedding works to increase the accuracy of detection. }
	\label{fig:fig10}
\end{figure}

Although this model has a relatively high degree of accuracy, there are several possibilities with which it can be improved. If one is solely interested in the locations of the nanoparticles, and other characteristics such as area and perimeter are of no value, the particles can simply be eroded to be separated and dilating process (which can cause nanoparticles to re-join) can be eliminated from the model. This method can further be improved by significantly eroding the first iteration to detect only the brightest particles and implementing more recursion loops to achieve a much higher degree of accuracy. Although more accurate, it would not be possible to characterize any other data from the particles using this method. 
A different method that could be pursued is a watershed algorithm. Watershed is an image segmentation transformation that is based on defining a true background and a true foreground that, when used with a distance transform, allow the watershed algorithm to detect boundaries between particles. The distance transform allows the algorithm to take the derivative of the binary image, finding maxima and minima particles. This is illustrated in Figure \ref{fig:fig10}.

These techniques allow the picture to become a 3-dimensional image, with hills and valleys. The watershed algorithm allows black pixels to ``flow'' around the image, separating objects that are clumped together. This technique holds the most promise for separating nanoparticles to achieve higher accuracy. Future research plans for this program include analysing SEM images of plasma nanoparticles deposited onto material using cathodic arc deposition in the lab.

\section{Conclusions}

This paper introduces a specialized computational framework designed for the precise detection and comprehensive analysis of nanoparticles within SEM images. The primary objective of accurately localizing nanoparticle coordinates was successfully achieved, along with the extraction of essential morphological attributes. Leveraging image processing techniques and libraries like OpenCV, SciPy, and Scikit-Image in Python, a robust pipeline was constructed to facilitate nanoparticle recognition.

The results demonstrate that the framework attains a remarkable level of accuracy in nanoparticle detection, achieving 97\% accuracy across multiple test images. The framework's recursive processing techniques excel in identifying faint nanoparticles that often evade manual labeling, showcasing its potential for heightened detection sensitivity. This paper not only addresses the initial research question concerning automating nanoparticle detection but also offers a solution that transcends the limitations of manual labeling, resulting in superior precision.

This study's applications extend beyond mere particle recognition, offering potential utility in diverse fields like cell biology, plasma physics, and material science. The technology streamlines data extraction, expedites analysis, and mitigates human error. This advancement obviates the necessity for labor-intensive and time-consuming manual labeling, making it accessible to users with varied levels of image processing expertise.

Looking ahead, this research lays the groundwork for further development and refinement. Future enhancements may involve the utilization of advanced techniques such as watershed algorithms to enhance particle separation accuracy. This avenue opens doors to diverse research directions, particularly in the analysis of various particle shapes and structures. By presenting a novel, recursive computational framework that addresses the challenges of nanoparticle detection within SEM images, this study successfully achieves high accuracy rates, demonstrates practical applications, and paves the way for automated particle recognition advancement in scientific research and analysis..

\end{document}